%% file: main.tex
\documentclass[a4paper]{article}
\usepackage[english]{babel}

\title{Fuzzy Arrovian Theorems when preferences are complete}
\author{Armajac Ravent\'os-Pujol}
\date{}

\input{packages}
\input{paragraphblocks}

\input{hardcode}

\begin{document}

\maketitle

\begin{abstract}
    In this paper we study the aggregation of fuzzy preferences on non-necessarily finite societies. We characterize in terms of possibility and impossibility a family of models of complete preferences in which the transitivity is defined for any t-norm.\\
    For that purpose, we have described each model by means of some crisp binary relations and we have applied the results obtained by Kirman and Sondermann.
\end{abstract}

\section{Introduction}

In the middle of the past century, Joseph Kenneth Arrow proved his impossibility theorem \cite{MR0039976, Arrow1963} modifying economists and social scientists' paradigm. His contribution triggered a considerable amount of research looking for functions which aggregate individual preferences in a single social preference representing the society as a whole.

From the Arrow's contribution, many alternative models have been developed (e.g. \cite{CHICHILNISKY1980165, 10.2307/1914083, SATTERTHWAITE1975187}). Some of them aggregate preferences in some specific situations, but no one has found a satisfactory model with universal pretensions as the Arrovian one.

In the eighties, the first articles applying fuzzy sets in the resolution of Arrow paradox were published \cite{Dutta1987215, Fung1975227}. The main problem addressed in this literature is the generalization of the axioms of the Arrovian model (which is grounded on (crisp) set theory) to fuzzy set theory and finding the aggregation functions that satisfy these new conditions in the fuzzy environment.\\

It turns out that there are too many possible generalizations of the Arovian model to the fuzzy environment. Many authors have discussed which types of generalizations fit better to the economic or behaviour sciences purposes and have considered some generalizations more suitable than others (see e.g. \cite{billot2012economic}).\\
Despite all this variety of models, we will focus on one of the most extended families of extensions (used e.g. in \cite{Banerjee1994121, Duddy201125, Gibilisco2014}). This family of models is characterized by a t-norm modelling the juxtaposition (intersection) and a t-conorm modelling the conjunction (union) in the fuzzy set framework.

In this framework, setting some combination of t-norms and t-conorms, there could appear some possibility results (e.g. \cite[Theorem 4.4]{arrowfuzzyA1} or \cite[Theorem 4.43]{Gibilisco2014}) and setting other combinations may lead to impossibility results (e.g. \cite[Proposition 3.5]{Banerjee1994121}). Moreover, there is are general results classifying all these models in terms of possibility and impossibility, but isolated results studying single combinations (e.g. \cite{Dutta1987215, Gibilisco2014}) or, exceptionally, a result about some subfamilies (e.g. \cite{Duddy201125}).\\

In this article, we will prove an impossibility result for a subfamily of Arrovian fuzzy models. We understand the study of this subfamily as an intermediate step towards the characterization of the whole family. In this subfamily of models, the t-conorm generalizing the union has no 1-divisors (i.e. complete preferences), whereas the t-norm generalizing the intersection can be anyone.

As we have said, we do not consider this work as a conclusion of our research in the field. Our goal is the classification of the whole family of models defined by t-norms and t-conorms. However, the following particularities of our are relevant per se and we consider that the scientific community may be interested in them:

First, we have been able to implement the technique we started in \cite{arrowfuzzyA1} and used in \cite{Raventos-Pujol202003}. This new technique is based on controlling fuzzy preferences by means of crisp preferences. It is important to remark that we developed this technique in models where the transitivity was not defined by a t-norm (for the definition of weak transitivity, see \cite[Definition 3.32]{Gibilisco2014}). So, applying the technique on the preferences defined by a t-norm requires a few adjustements. Besides, we must say that Billot in \cite{billot2012economic} remarks the deep difference between weak transitivity and transitivities defined by t-norms because of their contrast in terms of ordinality and cardinality. Fortunately, in this article, we have overcome this difference by adjusting the aforementioned technique to preferences in which a t-norm defines the transitivity instead of the weak transitivity.

Second, we have seen that complete fuzzy preferences have some exceptional properties. That is, we can use them as if they were a total preorder with a degree associated with every pair of elements. We use these properties as the cornerstone of our main theorem; however, they are interesting by themselves. We think that under this new light, they could acquire a new role in other models.\\

The article is structured as follows: After the introduction, we include a preliminaries section where we introduce the Arrovian fuzzy models. In section \ref{linearpreferences}, we analyze the properties of the complete preferences, and we discuss how they can be interpreted. In section \ref{aggregation}, we study the aforementioned subfamily of fuzzy Arrovian models and prove the corresponding theorems for non-necessarily finite societies, and we study finite societies as a particular case. Finally, there is a section of conclusions.

\section{Preliminaries}
\label{preliminaries}

In this paper, $X$ will denote a set containing 3 or more elements. Before introducing the fuzzy Arrovian models, we need to state some definitions from classic Set Theory.\\

A total preorder $\succsim$ in $X$ is a reflexive ($x\succsim x$ for every $x\in X$), transitive (if $x\succsim y$ and $y\succsim z$, then $x\succsim z$ for all $x,y,z\in X$) and complete ($x\succsim y$ or $y\succsim x$ for all $x,y\in X$) binary relation on $X$.\\
The asymmetric part $\succ$ of a preorder $\succsim$ is a binary relation in $X$ defined for every $x,y\in X$ as $x\succ y$ if $x\succsim y$ and not $y\succsim x$. The symmetric part $\sim$ of $\succsim$ is defined for every $x,y\in X$ as $x\sim y$ if $x\succsim y$ and $y\succsim x$. In particular, $\sim$ is an equivalence relation (reflexive, symmetric and transitive) and $\succ$ is an asymmetric and negatively transitive (if $x\succ z$, then $x\succ y$ or $y\succ z$ for all $x,y,z\in X$) binary relation (see \cite[Proposition 1.1.7]{Bridges1995}).\\
If $x\succsim y$, we say that $x$ is at least as good as $y$. Moreover, if $x\succ y$ (resp. $x\sim y$), we say that $x$ is preferred over $y$ (resp. equally preferred to). For that reason, we also name $\succsim$ as a weak preference relation, $\succ$ as a strict preference relation and $\sim$ as an indifference relation.

\begin{definition}A fuzzy preference in $X$ is a relation $R: X\times X \rightarrow [0,1]$. For every t-norm $T$ and t-conorm $S$\footnote{T-norms and t-conorms are operators widely used in fuzzy literature. See, for example, \cite[Chapter 3]{Beliakov2007} for the corresponding definitions} we say that:
\begin{itemize}
    \item [-] $R$ is reflexive if $R(x,x)=1$ for every $x\in X$,
    \item [-] $R$ is $T$-transitive if $R(x,z)\ge T\left(R(x,y),R(y,z)\right)$ for every $x,y,z\in X$,
    \item [-] $R$ is $S$-connected if $S\left(R(x,y),R(y,x)\right)=1$ for every $x,y\in X$.
\end{itemize}
If $S$ has no 1-divisors\footnote{A number $a\in (0,1)$ is a 1-divisor of $S$ if there exist a $b\in (0,1)$ such that $S(a,b)=1$ (see \cite[Definition 3.17]{Beliakov2007}).}, for every $x,y\in X$ $R(x,y)=1$ or $R(y,x)=1$ and we say that $R$ is complete.
\end{definition}

\begin{definition} Let $R$ be a relation in $X$ and $Y$ a subset of $X$. The restriction of $R$ in $Y$, $R_{\rceil Y}$,  is the fuzzy preference defined in $Y$ as $R_{\rceil Y}(x,y)=R(x,y)$ for every $x,y\in Y$. 
\end{definition}

A fuzzy preference $R$ plays the role of a binary relation $\succsim$ but in the fuzzy setting, we need the equivalent of the strict preference $\succ$ in the fuzzy setting. There are many definitions of the strict preference $P_R$ derived from $R$ (see \cite{Gibilisco2014, Ovchinnikov1981169, Raventos-Pujol202003}). However, in this article we do not require those amount of analysis and precision. So, we will suppose we have set a strict preference definition satisfying the following properties:
\begin{itemize}
    \item [i)] If $R(x,y)=1$ and $R(y,x)=0$, then $P_R(x,y)=1$,
    \item [ii)] $P_R(x,y)>0$ if, and only if, $R(x,y)>R(y,x)$,
    \item [ii)] $R(x,y)\ge R'(a,b)$ and $R(y,x)\le R'(b,a)$, then $P_R(x,y)\ge P_{R'}(a,b)$,
\end{itemize}
for every $x,y,a,b\in X$ and every pair of preferences $R$ and $R'$.\\

Let $N$ be a non-empty set representing a society of individuals. A profile of preferences for the society $N$ in a set of preferences $\mathcal{FP}$ is a function $\mathbf{R}: N \rightarrow \mathcal{FP}$. Besides, we denote the preference of the individual $i\in N$ as $R_i$ instead of $\mathbf{R}(i)$. If $N$ is finite, a profile is usually represented by the n-tuple $(R_1,\ldots,R_n)$ where $n=|N|$. We denote the set of all profiles in $\mathcal{FP}$ as $\mathcal{FP}^N$.

\begin{definition} Let $\mathcal{FP}$ be a set of fuzzy preferences. An aggregation function in $\mathcal{FP}$ is a function $f:\mathcal{FP}^N\rightarrow \mathcal{FP}$. We say that:
\begin{itemize}
    \item [-] $f$ is weak Paretian if, for every profile $\mathbf{R}\in\mathcal{FP}^N$ and pair of alternatives $x,y\in X$,  $P_{R_i}(x,y)>0$ for all $i\in N$ implies that $P_{f(\mathbf{R})}(x,y)>0$,
    \item [-] $f$ is strong Paretian if, for every profile $\mathbf{R}\in\mathcal{FP}^N$ and pair of alternatives $x,y\in X$, $P_{f(\mathbf{R})}(x,y)\ge \inf_{i\in N} P_{R_i}(x,y)$ holds,
    \item [-] $f$ satisfies independence of irrelevant alternatives (IIA) if, for every pair of profiles $\mathbf{R}, \mathbf{R}'\in\mathcal{FP}^N$ and pair of alternatives $x,y\in X$, if $R_{i\rceil\{x,y\}}=R'_{i\rceil\{x,y\}}$ for every $i\in N$ implies that $f(\mathbf{R})_{\rceil\{x,y\}}=f(\mathbf{R}')_{\rceil\{x,y\}}$,
    \item [-] $f$ is dictatorial if there is a $k\in N$ such that for every profile $\mathbf{R}\in\mathcal{FP}^N$ and pair of alternatives $x,y\in X$, $P_{R_k}(x,y)>0$ implies that $P_{f(\mathbf{R})}(x,y)>0$,
    \item [-] $f$ is strong dictatorial if there is a $k\in N$ such that for every profile $\mathbf{R}\in\mathcal{FP}^N$, pair of alternatives $x,y\in X$ and $\alpha\in [0,1)$, $P_{R_k}(x,y)>\alpha$ implies that $P_{f(\mathbf{R})}(x,y)>\alpha$.
\end{itemize}
\label{axioms}
\end{definition}

These are some of the generalizations of the axioms proposed in \cite{MR0039976} by Arrow to the fuzzy setting. They are not new in the literature (see e.g. \cite{billot2012economic, Dutta1987215, Gibilisco2014}). Perhaps, the least widespread is the strong dictator \cite{Banerjee1994121}.\\
In this paper, we will focus on the study of the aggregation functions in the sets of linear preferences defined below:

\begin{definition} Let $T$ be a t-norm. We say that a fuzzy preference in $X$ is $T$-linear if it is reflexive, complete and $T$-transitive. We denote the set of $T$-linear preferences by $\mathcal{LP}_T$.
\end{definition}

In the next sections, we will study the properties of the sets $\mathcal{LP}_T$, their preferences and how to aggregate them. Since the given arguments work for any t-norm, we set here any t-norm $T$, and we will denote $\mathcal{LP}_T$ by $\mathcal{LP}$ irrespective of the chosen t-norm $T$.

\section{Linear fuzzy preferences}
\label{linearpreferences}

A linear preference $R\in \mathcal{LP}$ is a special case of fuzzy preference. The fact that for every pair of alternatives $x,y\in X$ $R(x,y)=1$ or $R(y,x)=1$ holds (completeness), furnish $R$ with some properties that make it different from non-linear preferences.

This section is mainly devoted to prove that a linear preference is equivalent to a total preorder which has been enriched with extra structure. That is, each pair of alternatives $x,y\in X$ have a number (degree) from $[0,1]$ associated to them. In addition, the distribution of the degrees among the pairs of alternatives satisfy certain properties. First, we have to start with some definitions.

\begin{definition} For every preference $R\in \mathcal{LP}$, we associate to $R$ a binary relation $\succsim_R$ in $X$ defined as $x\succsim_R y$ if $R(x,y)=1$ (for every $x,y\in X$).
\end{definition}

\begin{proposition} For every $R\in \mathcal{LP}$, $\succsim_R$ is indeed a total preorder.
\begin{proof}
    It is immediate to check that $\succsim_R$ is reflexive as well as total. If $x\succsim_R y$ and $y\succsim_R z$, then $x\succsim_R z$ because $R(x,z)\ge T(R(x,y),R(y,z))=1$.
\end{proof}
\end{proposition}

\begin{remark} The associated preorder $\succsim_R$ of a linear preference $R$ motivates a new nomenclature. When we mention the qualitative behavior or properties of $R$, we are pointing to the properties of $\succsim_R$, whereas when we use the word quantitative, we are pointing to the degrees of $R$. For example, the fact $x\succsim_R y$ is qualitative, whereas $R(x,y)=0.3$ is quantitative. We will use these concepts to frame some propositions in the next section.
\end{remark}

The next result states an important relation between the degrees of a linear preference and its associated preorder.

\begin{lemma} Let $R\in \mathcal{LP}$ be a linear preference and $x,y,a,b\in X$ four alternatives. If $x\succsim_R a \succsim_R b\succsim_R y$, then $R(y,x)\le R(b,a)$.
\begin{proof} If $x\succsim_R a \succsim_R b\succsim_R y$, then $R(b,y)=R(x,a)=1$. We can deduce that $R(b,a)\ge T(R(b,y),R(y,a))=R(y,a)\ge T(R(y,x),R(x,a))=R(y,x)$.
\end{proof}
\label{graduacion}
\end{lemma}

The previous lemma has a nice interpretation. It says that if $a$ and $b$ are between $x$ and $y$, then $a$ and $b$ are more equivalent between them than $x$ and $y$. In the following example we illustrate how to visualize a linear preference as a total preorder with an extra structure. This schema has been used in most of the proofs.  

\begin{example} The schema below is a good representation of how we can imagine a linear preference $R$.\\

\begin{figure}[h]
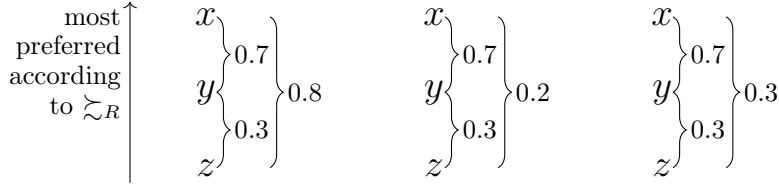

    \centering
    \include{esquema1}
    \caption{$R_1$ is not a linear preference whereas $R_2\in \mathcal{LP}_{T_{\text{\L}}}$ and $R_3\in \mathcal{LP}_{\min}$.}
\end{figure}

Here, the three alternatives are ranked as $x\succ_R y \succ_R z$. In the second case, $R_2(y,x)=0.7$, $R_2(z,y)=0.3$ and $R_2(z,x)=0.2$. Since $y$ is between $x$ and $z$, $R_2(y,x)\ge R_2(z,x)$. It can be interpreted as the degree of equivalence of $x$ and $y$ is greater than the degree of equivalence of $x$ and $z$. However, $R_1$ is not a linear preference because of the inequality $R_1(z,y)<T(R_1(z,x),R_1(x,y))=R_1(z,x)$.\\
In general, applying the inequality from Lemma \ref{graduacion} twice, we can deduce that $\min\{R(z,y),R(y,x)\}\ge R(z,x) \ge T(R(z,y),R(y,x))$. So, there is a constrain coming from being a linear preference and another from the t-norm itself. In particular, notice that $R_2$ does not belong to $\mathcal{LP}_{\min}$ but to $\mathcal{LP}_{T_{\text{\L}}}$\footnote{$T_{\text{\L}}$ denotes the \L ukasiewicz t-norm defined as $T_{\text{\L}}(x,y)=\max\{0,x+y-1\}$.}. From that, we deduce that the sets of linear preferences $\mathcal{LP}_T$ are different depending on the t-norm $T$.\\
Finally, for every $R\in\mathcal{LP}_{\min}$ $R(z,x)=\min\{R(z,y),R(y,x)\}$. So, we can state that the degrees between consecutive alternatives determine the whole preference $R$.
\end{example}

Finally we need a technical lemma that allows us to extend linear preferences in the same way that it is commonly done with total preorders. In many situation in classical Social Choice (e.g. \cite[Chapter 8]{kelly1988}), the reasoning is applied on three alternatives instead of the whole set $X$. These types of arguments are feasible because any preorder over a subset of $X$ can be trivially extended to the whole set $X$. The next lemma proves that we can make an equivalent extension in sets of linear preferences.

\begin{lemma} Let $Z \subseteq X$ and $\tilde{R}$ be a reflexive, $T$-transitive and complete preference defined on $Z$. Then, there is an extension of $\tilde{R}$. That is, a reflexive, $T$-transitive and complete preference $R$ satisfying $\tilde{R}=R_{\rceil Z}$.
\label{extension}
\begin{proof} Given $\tilde{R}$ defined on $Z$, define $R$ for every $\alpha , \alpha '\in Z$ and $\beta ,\beta '\in X\smallsetminus Z$ as $R(\alpha,\alpha ')=\tilde{R}(\alpha,\alpha ')$, $R(\alpha, \beta)=R(\beta, \beta ')=1$ and $R(\beta, \alpha)=0$.

It is clear that $R$ is reflexive and complete. It remains to see that it is $T$-transitive. 

If we suppose that $R$ is not $T$-transitive, there are $x,y,z\in X$ such that $R(x,z)<T(R(x,y),R(y,z))$. From this inequality we obtain that $R(x,z)<1$, $0<R(x,y)$ and $0<R(y,z)$. Using the definition of $R$ we deduce that: if $z\in Z$ then $y\in Z$, if $y\in Z$ then $x\in Z$ and if $x\in Z$, then $z\in Z$. In other words, the three alternatives belong or not to $Z$ together. This is a contradiction, because if they belong, they do not satisfy the inequality by hypothesis of $\tilde{R}$, but if they do not belong, $R(x,y)=R(y,z)=R(x,z)=1$ and the inequality is not satisfied.
\end{proof}
\end{lemma}

\begin{remark} Notice that the same proof can be used to non-complete fuzzy preferences since every complete preference is $S$-connected for any t-conorm $S$.
\end{remark}

Finally, we need the following definition for a technical purpose. Mainly, it will be useful when we need to reduce the degree of indifference to $0$ in order to obtain a linear preference with values in $\{0,1\}$. That is, a crisp preference or a total preorder.

\begin{definition} Let $R$ be a preference. We define $R^0$ as the fuzzy preference taking values in $\{0,1\}$ defined as $R^0(x,y)=1$ if and only if $R(x,y)=1$ and $R^0(x,y)=0$ otherwise. In the same way, given a profile $\mathbf{R}$, $\mathbf{R}^0$ is the profile defined as ${R^0}_i={R_i}^0$.
\end{definition}

\begin{definition} Let $\succsim$ be a total preorder on $X$. We define $R^\succsim$ as the fuzzy preference satisfying $R^\succsim(x,y)=1$ if $x\succsim y$ and $R^\succsim (x,y)=0$ otherwise.
\label{crisp}
\end{definition}

It is a routine to check that $R^\succsim$ and $R^0$ are linear preferences.

\begin{example} In the definition above, we have the most basic type of linear preferences. Another type of elemental linear preferences which will be used in this article are the ones whose degrees between alternatives are constant. That is, there is an $\alpha\in[0,1]$ and a total preorder $\succsim$ such that for every $x,y\in X$ $R(x,y)=1$ if $x \succsim y$ and $R(x,y)=\alpha$ if $y\succ_R x$.\\
In the case in which $\alpha=1$, the preference shows indifference between all alternatives, whereas when $\alpha=0$ we are in the situation of Definition \ref{crisp}.
\end{example}

The linear preferences explained in the previous example will play a central role in the next section. They, along with Lemma \ref{extension}, will facilitate the writing of the proofs.

\section{Aggregating linear preferences}
\label{aggregation}

This section will study the aggregation functions that satisfy the properties from Definition \ref{axioms}. We will follow a similar schema to the used in \cite{Hansson1976, Kirman1972267} for aggregating total preorders. However, those arguments can not be applied straightforwardly. After proving Propositions \ref{neutrality}, \ref{neutrality2} and Corollary \ref{ordinalIndependence}, we will be able to use ultrafilters in order to describe the aggregation function which we are interested in and prove the main results of this article.\\

The proposition below states that if the preference between a pair of alternatives is equal to another pair, then the social preference between these pairs have to coincide. This type of properties is usually named in the literature as neutrality. That is, the aggregation function is invariant with respect to permutations of alternatives.

\begin{proposition} Let $f$ be an aggregation function satisfying the independence of irrelevant alternatives and weakly Paretian. Then, for every pair of profiles $\mathbf{R}, \mathbf{R}'\in \mathcal{LP}^N$ and alternatives $a,b,x,y\in X$ the following holds: If for every $i\in N$ $R_i(x,y)=R'_i(a,b)$, $R_i(y,x)=R'_i(b,a)$ and $\min\{R_i(x,y),R_i(y,x)\}<1$, then $f(\mathbf{R})(x,y)=f(\mathbf{R}')(a,b)$ and $f(\mathbf{R})(y,x)=f(\mathbf{R}')(b,a)$.
\begin{proof} We can suppose without loss of generality that $x\succsim_{f(\mathbf{R})} y$. We define $\mathbf{R}^*$ using the Lemma \ref{extension} as the extension of the profile $\mathbf{\tilde{R}}^*$ defined on $\{x,y,a,b\}$ as $\tilde{R}^*_{i\rceil\{x,y\}}=R_{i\rceil\{x,y\}}$ and $\tilde{R}^*_{i\rceil\{a,b\}}=R'_{i\rceil\{a,b\}}$. Moreover, we define $\tilde{R}_i^*(a,x)=\tilde{R}_i^*(y,b)=1$, $\tilde{R}_i^*(x,a)=\tilde{R}_i^*(b,y)=a_i$ and if $x\succ_{R_i} y$ we define $\tilde{R}_i^*(a,y)=\tilde{R}_i^*(x,b)=1$ and $\tilde{R}_i^*(y,a)=\tilde{R}_i^*(b,x)=a_i$, on the contrary, if $y\succ_{R_i} x$ we define $\tilde{R}_i^*(a,y)=\tilde{R}_i^*(x,b)=a_i$ and $\tilde{R}_i^*(y,a)=\tilde{R}_i^*(b,x)=1$. Using independence of irrelevant alternatives and weak Pareto condition, we obtain that $a \succ_{f(\mathbf{R}^*)} x\succsim_{f(\mathbf{R}^*)} y\succ_{f(\mathbf{R}^*)} b$. Moreover, if we apply Lemma \ref{graduacion}, we obtain that $f(\mathbf{R}^*)(y,x)\ge f(\mathbf{R}^*)(b,a)$. Finally, using  independence of irrelevant alternatives, we obtain that $f(\mathbf{R})(y,x)\ge f(\mathbf{R})(b,a)$\\
We can use a similar argument in order to prove that $f(\mathbf{R})(y,x)\le f(\mathbf{R})(b,a)$. It is only necessary to define $R_i^*(x,a)=R_i^*(b,y)=1$, $R_i^*(a,x)=R_i^*(y,b)=a_i$.
\end{proof}
\label{neutrality}
\end{proposition}

The next proposition and corollary state that if the qualitative behaviour of two profiles coincide, then its aggregation also have to coincide.  

\begin{proposition} Let $f$ be an aggregation function satisfying the  independence of irrelevant alternatives and weakly Paretian, and $x,y\in X$. If for every $i\in N$ a profile $\mathbf{R}$ satisfies $x \succ_{R_i} y$ or $y\succ_{R_i} x$, then $\succsim_{f(\mathbf{R})\rceil\{x,y\}}=\succsim_{f(\mathbf{R}^0)\rceil\{x,y\}}$.
\begin{proof} Set a profile $\mathbf{R}$ satisfying the conditions of the proposition. We can suppose without loss of generality that $x\succsim_{f(\mathbf{R})} y$. We choose a third alternative $z\in X$ and we define a profile $\mathbf{R}'$ satisfying for every $i\in N$ $R_{i\rceil\{x,y\}}=R'_{i\{x,y\}}$, $R'_i(z,y)=R_i(x,y)$, $R'_i(y,z)=R_i(y,x)$ and defined for the other pairs according to Lemma \ref{extension}. Proposition \ref{neutrality} guarantees that $z\succsim_{f(\mathbf{R}')} y$. We define the profile $\mathbf{R}''$ for every $i\in N$ $R''_{i\rceil\{x,y\}}=R^0_{i\rceil\{x,y\}}$, $R''_{i\rceil\{z,y\}}=R^0_{i\rceil\{z,y\}}$, $R''_i(z,x)=0$, $R''_i(x,z)=1$ and the remaining pairs according to Lemma \ref{extension}. Applying the independence of irrelevant alternatives, we obtain that $z\succsim_{f(\mathbf{R})} y$. Applying the weak Paretian property we obtain that $x\succ_{f(\mathbf{R}'')} z$, and we obtain $x\succ_{f(\mathbf{R}'')} y$. Applying the independence of irrelevant alternatives again, we obtain that $x\succ_{f(\mathbf{R}^0)} y$.\\
Moreover, notice that $x \sim_{f(\mathbf{R})} y$ is not possible. If it were the case, we could apply again the same procedure as above but starting from $y\succsim_{f(\mathbf{R})} x$. In that way, we would obtain that $y\succ_{f(\mathbf{R}^0)} x$ too. However, it is a contradiction. We state that $x\succ_{f(\mathbf{R})} y$ or $y\succ_{f(\mathbf{R})} x$, and this way we can conclude that $\succsim_{f(\mathbf{R})\rceil\{x,y\}}=\succsim_{f(\mathbf{R}^0)\rceil\{x,y\}}$.
\end{proof}
\label{neutrality2}
\end{proposition}

\begin{corollary} Let $f$ be an aggregation function satisfying the independence of irrelevant alternatives and weakly Paretian. For every pair of profiles $\mathbf{R},\mathbf{R}'$ and alternatives $x,y\in X$ such that $\succsim_{R_i\rceil\{x,y\}}=\succsim_{R_i\rceil\{x,y\}}$ and $x\not\sim_{R_i} y$ for every $i\in N$, then  $\succsim_{f(\mathbf{R})\rceil\{x,y\}}= \succsim_{f(\mathbf{R}')\rceil\{x,y\}}$.
\begin{proof} If $\mathbf{R}$ and $\mathbf{R}'$ satisfies the previous conditions, for every $i\in N$, ${{R_i}^0}_{\rceil\{x,y\}}={{R'_i}^0}_{\rceil\{x,y\}}$ holds. Then, applying the previous proposition, we obtain that\\ $\succsim_{f(\mathbf{R})\rceil\{x,y\}}=\succsim_{{f(\mathbf{R}}^0)\rceil\{x,y\}}=\succsim_{f(\mathbf{R}')\rceil\{x,y\}}$.
\end{proof}
\label{ordinalIndependence}
\end{corollary}

Finally, we can proceed studying the aggregation functions using ultrafilters. First, we need to recall what is a filter and an ultrafilter:

\begin{definition} Let $A$ be a set and $\mathfrak{F}$ a family of subsets of $A$. We say that $\mathfrak{F}$ is a filter if for every $U,V\subseteq A$ the following conditions are satisfied:
\begin{itemize}
    \item [i)] $\emptyset \notin \mathfrak{F}$,
    \item [ii)] if $U\in \mathfrak{F}$ and $U\subseteq V$, then $V\in \mathfrak{F}$,
    \item [iii)] if $U,V\in \mathfrak{F}$, then $U\cap V \in \mathfrak{F}$.
\end{itemize}
Moreover, we say that a filter $\mathfrak{U}$ is an ultrafilter if it is a maximal filter, that is, if for every filter $\mathfrak{F}$ satisfying $\mathfrak{U}\subseteq \mathfrak{F}$, we obtain that $\mathfrak{U}=\mathfrak{F}$.
\end{definition}

\begin{proposition} Let $\mathfrak{F}$ be a filter on a set $A$. $\mathfrak{F}$ is an ultrafilter if and only if for every $V\subseteq A$, $V\in \mathfrak{F}$ or $V^c\in \mathfrak{F}$.
\begin{proof} See, for example, \cite[Theorem 12.11]{willardtopology}.
\end{proof}
\end{proposition}

We will see that the decisive coalitions with respect to an aggregation functions define, in fact, an ultrafilter on the society. First, we have to give a formal definition for the decisive coalitions:

\begin{definition} Let $f$ an aggregation function in $\mathcal{LP}$ in the society $N$. A coalition $C\subseteq N$ is decisive if for every pair $x,y\in X$ and profile $\mathbf{R}\in \mathcal{LP}^N$ the following condition is satisfied:
\begin{equation*}
    \left[\forall i\in C x\succ_{R_i} y \text{ and } \forall i\in C^c y\succ_{R_i} x\right] \Rightarrow x\succ_{f(\mathbf{R})} y
\end{equation*}
\end{definition}

We denote the set of all decisive coalitions of $f$ as $\mathcal{D}_f$.\\

Now, we have all the requirements to prove that $\mathcal{D}_f$ is an ultrafilter. As we have said, the arguments below are inspired by the ones used by Kirman and Sonderman in \cite{Kirman1972267}. There, they proved that the set of coalitions for a crisp aggregation function is an ultrafilter. If we do the exercise of comparing the proofs of the remaining parts of this section with the ones in \cite{Kirman1972267}, we will see that the previous propositions are the key to the adaptations we have made in the proofs.\\

First, the next proposition provides us with three equivalent definitions of decisive coalitions.

\begin{proposition} Let $f$ be an aggregation function satisfying the independence of irrelevant alternatives as well as weak Pareto property. Then the following three sets are equal:
\begin{itemize}
    \item [] $\mathcal{D}''_f=\{C\subseteq N: \exists\;x,y\in X\; \; \exists \mathbf{R} \; x \succ_{R_C} y \text{ and } y\succ_{R_{C^c}} x \text{ and } x\succ_{f(\mathbf{R})} y\}$
    \item [] $\mathcal{D}'_f=\{C\subseteq N: \exists\;x,y\in X\; \; \forall \mathbf{R} \; x \succ_{R_C} y \text{ and } y\succ_{R_{C^c}} x \Rightarrow x\succ_{f(\mathbf{R})} y\}$
    \item [] $\mathcal{D}_f=\{C\subseteq N: \forall\;x,y\in X\; \; \forall \mathbf{R} \; x \succ_{R_C} y \text{ and } y\succ_{R_{C^c}} x \Rightarrow x\succ_{f(\mathbf{R})} y\}$
\end{itemize}
\begin{proof} It is straightforward to check that $\mathcal{D}_f\subseteq \mathcal{D}'_f\subseteq \mathcal{D}''_f$.
To prove that $\mathcal{D}''_f\subseteq \mathcal{D}'_f$, using Corollary \ref{ordinalIndependence} is enough.
Finally, to prove that $\mathcal{D}'_f\subseteq \mathcal{D}_f$, we use Proposition \ref{neutrality}.
\end{proof}
\end{proposition}

The next proposition shows, finally, that the set of decisive coalitions is an ultrafilter.

\begin{proposition} Let $f$ be an aggregation function satisfying the  independence of irrelevant alternatives as well as weak Pareto property. Then $\mathcal{D}_f$ is an ultrafilter.
\begin{proof} First, since $f$ is weakly Paretian, $\emptyset \notin \mathcal{D}_f$.\\
Secondly, if $U,W\in \mathcal{D}_f$ and we want to prove that $U\cap W\in \mathcal{D}_f$, we set three different alternatives $x,y,z\in X$ and define a profile $\mathbf{R}$ as:
\begin{itemize}
    \item [-] for every $i\in U\cap W$, $R_i(x,z)=R_i(x,y)=R_i(z,y)=1$ and $R_i(z,x)=R_i(y,x)=R_i(y,z)=0$ ($x\succ_{R_i} z \succ_{R_i} y$ holds),
    \item [-] for every $i\in U\smallsetminus W$, $R_i(z,x)=R_i(z,y)=R_i(y,x)=1$ and $R_i(z,x)=R_i(y,z)=R_i(x,y)=0$ ($z\succ_{R_i} y \succ_{R_i} x$ holds),
    \item [-] for every $i\in W\smallsetminus U$, $R_i(y,x)=R_i(y,z)=R_i(x,z)=1$ and $R_i(x,y)=R_i(z,y)=R_i(z,x)=0$ ($y\succ_{R_i} x \succ_{R_i} z$ holds),
    \item [-] for every $i\notin W \cup U$, $R_i(y,x)=R_i(y,z)=R_i(z,x)=1$ and $R_i(x,y)=R_i(z,y)=R_i(x,z)=0$ ($y\succ_{R_i} z \succ_{R_i} x$ holds).
\end{itemize}
We complete the definition of $\mathbf{R}$ over the remaining pairs of alternatives as in Lemma \ref{extension}.\\
Since $U\in \mathcal{D}_f$, we obtain that $z\succ_{f(\mathbf{R})} y$ and from $W\in \mathcal{D}_f$ we obtain $x\succ_{f(\mathbf{R})} y$. Using the transitivity of $\succ_{f(\mathbf{R})}$, we obtain that $x\succ_{f(\mathbf{R})} y$. This proves that $U\cap W\in \mathcal{D}''_f=\mathcal{D}_f$.\\
Now, we will prove that for every $U\subseteq N$, $U\in \mathcal{D}_f$ or $U^c\in \mathcal{D}_f$ holds. We set an alternative $z\in X$ and a total preorder $\succsim$ in $X\smallsetminus \{z\}$ without indifferences. We define a profile $\mathbf{R}$ as:
\begin{itemize}
    \item [-] for every $i\in N$, $R_{i\rceil X\smallsetminus \{z\}}=R^{\succsim}$,
    \item [-] for every $i\in U$ and every $s\in X\smallsetminus\{z\}$, $R_i(s,z)=1$ and $R_i(z,s)=0$,
    \item [-] for every $i\in U^c$ and every $s\in X\smallsetminus\{z\}$, $R_i(z,s)=1$ and $R_i(s,z)=0$.
\end{itemize}
Notice that this profile is well defined because it is crisp. Set two alternatives $x,y\in X\smallsetminus\{z\}$. We can suppose without loss of generality that $x\succ y$, and by weak Pareto property we state that $x\succ_{f(\mathbf{R})} y$. If we use that $\succ$ is negatively transitive, we obtain that $x\succ_{f(\mathbf{R})} z$ or $z\succ_{f(\mathbf{R})} y$. Then $U\in \mathcal{D}''_f=\mathcal{D}_f$ or $U^c\in \mathcal{D}''_f=\mathcal{D}_f$.\\
Finally, we suppose that $U\in \mathcal{D}_f$ and $U\subseteq W$. If $W\notin \mathcal{D}_f$, then $W^c\in \mathcal{D}_f$, and this implies that $\emptyset= U\cap W^c\in \mathcal{D}_f$. However, at the beginning of the proof we have proved that this is not possible. We conclude that $W\in \mathcal{D}_f$.
\end{proof}
\label{isUltrafilter}
\end{proposition}

We consolidate the proposition with the following theorem. It shows that we can assign a unique ultrafilter to every aggregation function. 

\begin{theorem} Let $f$ be an aggregation function in the society $N$. If $f$ is weakly Paretian and satisfies the  independence of irrelevant alternatives, then there is a unique ultrafilter $\mathfrak{U}$ such that for every profile $\mathbf{R}\in \mathcal{LP}^N$ and every pair of alternatives $x,y\in X$, the following holds for every coalition $U\in \mathfrak{U}$:
\begin{equation*}
    \text{if for every }i\in U \; x\succ_{R_i} y\text{, then } x\succ_{f(\mathbf{R})} y.
\end{equation*}
Moreover, if $f$ is strongly Paretian, the following inequality is satisfied:
\begin{equation*}
    P_{f(\mathbf{R})}(x,y)\ge \inf_{i\in U} P_{R_i}(x,y).
\end{equation*}
\begin{proof} Using Proposition \ref{isUltrafilter} we obtain that $\mathcal{D}_f$ is an ultrafilter. First, we will check that it satisfies the statements of this theorem and later that it is unique.\\
Given a $U\in \mathcal{D}_f$, we consider a profile $\mathbf{R}$ and a pair of alternatives $x,y\in X$ with $x\succ_{R_i} y$ for every $i\in U$. We define a partition of $N=N_1\cup N_2 \cup N_3$ as $N_1=\{i\in N: x\succ_{R_i} y\}$, $N_2=\{i\in N: y\succsim_{R_i} x\}$ and $N_3=\{i\in N: x\sim_{R_i} y\}$. After, we set a third alternative $z\in X$ and we define another profile $\mathbf{R}'$ as:
\begin{itemize}
    \item [-] $R'_i(x,y)=R'_i(x,z)=R'_i(z,y)=1$ and $R'_i(y,x)=R'_i(z,x)=R'_i(y,z)=R_i(y,x)$ for every $i\in N_1$,
    \item [-] $R'_i(y,x)=R'_i(z,x)=R'_i(y,z)=1$ and $R'_i(x,y)=R'_i(x,z)=R'_i(z,y)=R_i(x,y)$ for every $i\in N_1$,
    \item [-] $R'_i(x,y)=R_i'(y,x)=R_i'(x,z)=R'_i(y,z)=1$ and $R'_i(z,x)=R'_i(z,y)=0$ for every $i\in N_3$
\end{itemize}
and defined over the remaining pairs as in Lemma \ref{extension}. Since $U\subseteq N_1$, $N_1\in \mathcal{D}_f$. Using the decisivity of $N_1$, we conclude that $z\succ_{f(\mathbf{R}')} y$. Applying the same argument on $N_1\cup N_3$, we obtain that $x\succ_{f(\mathbf{R}')} z$. We can conclude that $x\succ_{f(\mathbf{R}')} y$, and applying the  independence of irrelevant alternatives we obtain that $x\succ_{f(\mathbf{R})} y$.\\
To prove the uniqueness, consider two different ultrafilters $\mathfrak{U}$ and $\mathfrak{U}'$ satisfying the conditions of the theorem. Then, there exist a coalition $U\in \mathfrak{U}\smallsetminus \mathfrak{U}'$. Then, we can see that $U^c\in \mathfrak{U}'$. We can consider a profile $\mathbf{R}$ and a pair of alternatives $x,y\in X$ satisfying $x\succ_{R_U} y$ and $y\succ_{R_{U^c}} x$. However, since both of them satisfy the conditions of the theorem, we obtain that $x\succ_{f(\mathbf{R})} y \succ_{f(\mathbf{R})} x$, and it is a contradiction. We can conclude that the ultrafilter of the theorem is unique.\\
Finally, in order to prove the last inequality, it is enough to prove that if $x\succ_{f(\mathbf{R})} y$, then $f(\mathbf{R})(y,x)\le \sup_{i\in U} R_i (y,x)$. First, we can suppose that for every $i\in U$ $x\succ_{R_i} y$. Set an alternative $z\in X$ and define the profile $\mathbf{R}'$ as:
\begin{itemize}
    \item [-] $R'_{i\rceil\{x,y\}}=R_{i\rceil\{x,y\}}$,
    \item [-] $R'_i(x,z)=R'_i(z,y)=1$ and $R'_i(z,x)=R'_i(y,z)=R_i(y,x)$ if $i\in U$,
    \item [-] $R'_i(x,z)=R(y,z)=1$ and $R'_i(z,x)=R'_i(z,y)=0$ if $i\notin U$.
\end{itemize} 
and the image over the remaining pair of alternatives as in Lemma \ref{extension}. First, since $U$ is decisive, we obtain that $x\succ_{f(\mathbf{R})} z \succ_{f(\mathbf{R})} y$. If we apply Lemma \ref{graduacion} we obtain that $f(\mathbf{R})(y,x)\le f(\mathbf{R})(z,x)$. Hence, if we apply the strong Pareto criterion, we obtain that $f(\mathbf{R})(z,x)\le \sup_{i\in N} R_i(z,x)=\sup_{i\in U} R_i(z,x)=\sup_{i\in U} R_i(y,x)$.
\end{proof}
\label{Qimposibility}
\end{theorem}

The theorem below studies the opposite direction. It shows that every ultrafilter comes from an aggregation function. However, we do not have any insight about the uniqueness. That is, different aggregation functions may induce the same ultrafilter.

\begin{theorem} Let $N$ be a society and $\mathfrak{U}$ an ultrafilter in $N$. There is an aggregation function $f$ satisfying the independence of irrelevant alternatives and strong Paretian whose set of decisive coalitions is $\mathcal{D}_f=\mathfrak{U}$.
\begin{proof}
Set an ultrafilter $\mathfrak{U}$ and define the aggregation function $f$ as:
\begin{equation*}
    f(\mathbf{R})(x,y)=\begin{cases}
            1 &\text{if } \{i\in N: x\succ_{R_i} y\}\in \mathfrak{U},\\
            0 &\text{if }  \{i\in N: y\succ_{R_i} x\}\in \mathfrak{U},\\
            1 &\text{otherwise}.
        \end{cases}
\end{equation*}
It is straightforward to check that $f(\mathbf{R})$ is reflexive and complete for every $\mathbf{R}\in \mathcal{LP}^N$. If $f(\mathbf{R})$ were not transitive, then there would be three alternatives $x,y,z\in X$ such that $f(\mathbf{R})(x,y)<T\left(f(\mathbf{R})(x,z),f(\mathbf{R})(z,y)\right)$, so $f(\mathbf{R})(x,y)=0$ and $f(\mathbf{R})(x,z)=f(\mathbf{R})(z,y)=1$. Then $\{i\in N: y\succ_{R_i} x\} \in \mathfrak{U}$ and $\{i\in N: z\succ_{R_i} x\},\{i\in N: y\succ_{R_i} z\}\notin \mathfrak{U}$, and we can state that $\{i\in N: z\precsim_{R_i} x\},\{i\in N: y\precsim_{R_i} z\}\in \mathfrak{U}$. Finally, we can conclude that $\{i\in N: y\precsim_{R_i} x\}\in \mathfrak{U}$ because $\{i\in N: y\precsim_{R_i} x\}\supseteq\{i\in N: z\precsim_{R_i} x\}\cap\{i\in N: y\precsim_{R_i} z\}$. This is a contradiction because $\{i\in N: y\precsim_{R_i} x\}=\{i\in N: y\succ_{R_i} x\}^c$.\\
Finally, to prove that $\mathfrak{U}=\mathcal{D}_f$, it is enough to see that $\mathfrak{U}\subseteq \mathcal{D}''_f$. Given an $U\in \mathfrak{U}$, consider a profile $\mathbf{R}$ and a pair of alternatives $x,y\in X$ such that $x\succ_{R_i} y$ if $i\in U$ and $y \succ_{R_i} x$ otherwise. It is clear that $1=f(\mathbf{R})(x,y)>f(\mathbf{R})(y,x)=0$. Then $U\in \mathcal{D}''_f=\mathcal{D}_f$.
\end{proof}
\label{existenceFromU}
\end{theorem}

In most of the literature, the results of aggregation are applied over finite societies. Next, we will see the consequences of the previous theorems in finite societies.

\begin{corollary} Let be $f$ an aggregation function on $\mathcal{LP}$ and $N$ finite. If $f$ is weakly Paretian and satisfies the independence of irrelevant alternatives, then $f$ is dictatorial. Besides, if $f$ is strongly Paretian, $f$ is strong-dictatorial.
\begin{proof} If $N$ is finite, every ultrafilter $\mathfrak{U}$ in $N$ is generated by a single element, that is, there is an $i\in N$ that $\mathfrak{U}=\{V\subseteq N: i\in V\}$\footnote{It is easy to check from the definition: If it were not the case, then for all $j\in N$ $\{j\}\notin \mathfrak{U}$. Then, for all of them, $\{j\}^c\in \mathfrak{U}$. However, since $N$ is finite, $\emptyset=\bigcap_{j\in N} \{j\}^c \in \mathfrak{U}$. But this contradicts the definition of a filter.}. By Theorem \ref{Qimposibility}, $\mathcal{D}_f$ is an ultrafilter in $N$, then $\mathcal{D}_f=\{k\}$ for some $k\in N$. Clearly, $k$ is a dictator.\\
Moreover, if $f$ is strongly Paretian, by the same theorem, $P_{f(\mathbf{R})}\ge P_{R_k}$ for every profile $\mathbf{R}\in \mathcal{LP}$, this implies that $k$ is a strong dictator.
\end{proof}
\end{corollary}

\section{Conclusions}

The main achivements of this paper are the Theorem \ref{Qimposibility} characterizing the imposibility of a family of Arrovian models as well as the propositions that allows us to use arguments from Kirman and Sonderman paper in these fuzzy models.\\
In this work, we have described a new fuzzy model by means of crisp binary relations similarly as it has been done in other works with the same spirit (e.g. \cite{BASILE201899, axioms6040029, Raventos-Pujol202003}). We decided to focus on complete preferences instead of general $S$-connected preferences as a starting point of a more general study. Indeed, complete preferences have allowed us to define an associated preorder and use techniques from crisp literature. A priory, the same arguments can not be applied over more general sets of preferences.\\
Propositions \ref{neutrality}, \ref{neutrality2} and Corollary \ref{ordinalIndependence} show that the completeness condition is really strong. First, these intermediate results are the keystone of Theorem \ref{Qimposibility} in which the model's impossibility is proved. However, we can deduce two additional properties of the model; one of them is proper from classical Social Choice, and the second one from fuzzy modelization.\\
The first one is the neutrality; that is, the aggregation function has a symmetry with respect to the alternatives. Although neutrality is not imposed as an axiom in the standard fuzzy literature, most of the aggregation fuzzy functions are neutral (e.g. \cite[Lemma 4]{Duddy201125}, \cite[Proposition 3.9]{Dutta1987215} or \cite[Theorem 4.43]{Gibilisco2014}). This fact raises the question of whether we could derive the neutrality from the axioms in most of the cases as it happens in the Arrovian model or in Proposition \ref{neutrality}.\\
Second, Proposition \ref{neutrality2} proves that two profiles with the same qualitative behaviour have the same aggregation in qualitative terms. This is quite surprising because the independence of irrelevant alternatives is defined in quantitative terms. Notice that other models in the literature use qualitative formulations of the independence of irrelevant alternatives property (see \cite{billot2012economic, Mordeson2012219}). In these models, it is quite natural to build associated preorders compatible with the IIA property and use them to describe the aggregation function (as we made in \cite{Raventos-Pujol202003}). However, in this paper, the qualitative IIA is a consequence of the quantitative IIA and the completeness. It will be interesting to study the relations between the extensions of IIA deeply. As far as we know, there is not extended studies about the relations between distinct IIA properties.\\

We have studied the role of strong dictators. This type of dictators appears in some papers (e.g. \cite{Banerjee1994121}), but their study is not widespread. We should ask ourselves the reason behind the importance of being strictly greater than 0 ($P_R(x,y)>0$) and the lack of importance of being grater than other numbers ($P_R(x,y)>\alpha$, $\alpha \in [0,1)$). In a fuzzy model, where the concept of vagueness is the central point, which is the difference between, for example, the degree $\alpha = 0$ and the degree $\alpha = 10^{-50}$?\\

Finally, we need to make some comments about Theorem \ref{existenceFromU}. This theorem shows that every ultrafilter has at least one associated aggregation function, but it is not necessarily unique. In the proof, we define an aggregation function whose image only contains preferences with values in $\{0,1\}$; in other words, they are crisp functions. However, other aggregation rules which take into account intermediate degrees may exist.\\
In addition, there is a little confusion in some papers in the literature about the qualitative behaviour of the rules defined using ultrafilters. For instance, consider an ultrafilter $\mathfrak{U}\subseteq \mathcal{P}(N)$. Given a pair of alternatives $x,y\in X$ and a profile $\mathbf{R}\in \mathcal{LP}$, since we expect to be $\mathfrak{U}=\mathcal{D}_f$, if $\{i\in N: x\succ_{R_i} y\}\in \mathfrak{U}$ has to imply that $x\succ_{f(\mathbf{R})} y$. However, when $\{i\in N: x\succ_{R_i} y\}, \{i\in N: y\succ_{R_i} x\}\notin \mathfrak{U}$\footnote{It can be proved that this condition is equivalent to say that $\{i\in N: x\sim_{R_i} y\}\in \mathfrak{U}.$}, then we can define the image of $f$ without any constraint coming from $\mathfrak{U}$, that is, the three cases $x\succ_{f(\mathbf{R})} y$, $y\succ_{f(\mathbf{R})} x$ or $x\sim_{f(\mathbf{R})} y$ may be feasible. In Theorem \ref{existenceFromU}, the third option (i.e. $f(\mathbf{R})(x,y)=f(\mathbf{R})(y,x)=1$) have been chosen for all situations of indeterminacy.\\
If we were to characterize all aggregation functions compatible with a given ultrafilter, we would have to consider the quantitative and the qualitative indeterminacies explained above to create a good classification.

\section{Future Research}
In the future, we will study the sets of non-complete $S$-connected preferences. We will try to extrapolate the same technique used in this article to a more general case. For this purpose, we think about using other types of order binary relations (for instance, quasi-transitive binary relations or interval orders), or we could associate a family of binary relations to every fuzzy preference instead of a single one.\\
Using one of these adjustments, we could try to describe the qualitative behaviour of fuzzy preferences as we have made in this article using total preorders.

\bibliographystyle{abbrv} 
\nocite{*}
\bibliography{main}

\end{document}

%% file: packages.tex
\usepackage[utf8]{inputenc} 
\usepackage[T1]{fontenc}

\usepackage[colorlinks=true,
    linkcolor=blue,
    filecolor=magenta,      
    urlcolor=cyan,
    citecolor=blue,
    bookmarks=true]{hyperref}

\usepackage{fancyhdr}

\usepackage{amsfonts,amssymb,amsthm,amsmath}
\usepackage{xfrac}    
\usepackage{faktor}

\usepackage[dvipsnames]{xcolor}

\usepackage{multicol}
\usepackage{newunicodechar}

\usepackage{graphicx}
\usepackage{float}

\usepackage{bibentry}

\usepackage{tikz-cd}
\usepackage{pgf,tikz}

\usetikzlibrary{arrows}
\usetikzlibrary{patterns,intersections}
\usetikzlibrary{math}
\usetikzlibrary{decorations.pathreplacing,angles,quotes}
\usetikzlibrary{decorations.markings}
\usetikzlibrary{positioning}

\usetikzlibrary{patterns}

\usepackage{standalone}

\usepackage[toc]{glossaries}
\usepackage{glossary-mcols}

\usepackage{lscape}
\usepackage{subcaption}

%% file: paragraphblocks.tex
	\fancypagestyle{plain}{%first page of the chapter
		\fancyhf{}

	}

\numberwithin{equation}{section}
\theoremstyle{plain}
\newtheorem{theorem}{Theorem}[section]

\newtheorem{proposition}[theorem]{Proposition}

\newtheorem{corollary}[theorem]{Corollary}

\newtheorem{lemma}[theorem]{Lemma}

\theoremstyle{definition}
\newtheorem{definition}[theorem]{Definition}

\newtheorem{example}[theorem]{Example}

\theoremstyle{remark}
\newtheorem{remark}[theorem]{Remark}

%% file: hardcode.tex
\newunicodechar{Σ}{\Sigma}
\newunicodechar{φ}{\phi}
\newunicodechar{λ}{\lambda}
\newunicodechar{Λ}{\Lambda}
\newunicodechar{δ}{\delta}
\newunicodechar{θ}{\theta}
\newunicodechar{Ω}{\Omega}
\newunicodechar{μ}{\mu}
\newunicodechar{η}{\eta}
\newunicodechar{κ}{\kappa}
\newunicodechar{ν}{\nu}
\newunicodechar{ω}{\omega}
\newunicodechar{Φ}{\Phi}
\newunicodechar{Ξ}{\Xi}
\newunicodechar{α}{\alpha}
\newunicodechar{γ}{\gamma}
\newunicodechar{β}{\beta}
\newunicodechar{ς}{\varsigma}
\newunicodechar{π}{\pi}
\newunicodechar{Π}{\Pi}
\newunicodechar{σ}{\sigma}
\newunicodechar{Γ}{\Gamma}
\newunicodechar{ρ}{\rho}
\newunicodechar{ι}{\iota}
\newunicodechar{ε}{\epsilon}
\newunicodechar{Δ}{\Delta}
\newunicodechar{τ}{\tau}
\newunicodechar{χ}{\chi}

%% file: esquema1.tex
\begin{tikzpicture}[scale=1, every node/.style={scale=1}]
\tikzstyle{every node}=[font=\Large]
%\draw[help lines,gray, thin] (-10,-10) grid (10,15);

\tikzmath{\x=1; \y=0; \z=-1; \e=0.2;
            \a=3; \b=6;}

\node[name path=X] at (0,\x) {$x$};
\node[name path=Y] at (0,\y) {$y$};
\node[name path=Z] at (0,\z) {$z$};

%\draw [thick] plot [smooth, tension=2] coordinates {(\e,\x) (2*\e,\x*0.5+\y*0.5) (\e,\y)};
%\draw [thick] plot [smooth, tension=2] coordinates {(\e,\y) (2*\e,\z*0.5+\y*0.5) (\e,\z)};

%\node[name path=Y] at (2*\e+\e,\x*0.5+\y*0.5) {\normalsize $0.7$};
%\node[name path=X] at (2*\e+\e,\z*0.5+\y*0.5) {\normalsize $0.3$};

\draw [decorate,decoration={brace,amplitude=5pt,raise=4pt},yshift=0pt, xshift=0.7cm]
(0,\x) -- (0,\z) node [black,midway,xshift=0.6cm] {\normalsize
$0.8$};
\draw [decorate,decoration={brace,amplitude=5pt,raise=4pt},yshift=0pt, xshift=0cm]
(0,\x) -- (0,\y) node [black,midway,xshift=0.6cm] {\normalsize
$0.7$};
\draw [decorate,decoration={brace,amplitude=5pt,raise=4pt},yshift=0pt, xshift=0cm]
(0,\y) -- (0,\z) node [black,midway,xshift=0.6cm] {\normalsize
$0.3$};

%%%%%%%%%%%%%%%%%%%%%%%%%%%%%%%%%%%%%%%%%%%%%%%%%%%%%%

\node[name path=X] at (\a,\x) {$x$};
\node[name path=Y] at (\a,\y) {$y$};
\node[name path=Z] at (\a,\z) {$z$};

\draw [decorate,decoration={brace,amplitude=5pt,raise=4pt},yshift=0pt, xshift=0.7cm]
(\a,\x) -- (\a,\z) node [black,midway,xshift=0.6cm] {\normalsize
$0.2$};
\draw [decorate,decoration={brace,amplitude=5pt,raise=4pt},yshift=0pt, xshift=0cm]
(\a,\x) -- (\a,\y) node [black,midway,xshift=0.6cm] {\normalsize
$0.7$};
\draw [decorate,decoration={brace,amplitude=5pt,raise=4pt},yshift=0pt, xshift=0cm]
(\a,\y) -- (\a,\z) node [black,midway,xshift=0.6cm] {\normalsize
$0.3$};

%%%%%%%%%%%%%%%%%%%%%%%%%%%%%%%%%%%%%%%%%%%%%%%%%%%%%%%

\node[name path=X] at (\b,\x) {$x$};
\node[name path=Y] at (\b,\y) {$y$};
\node[name path=Z] at (\b,\z) {$z$};

\draw [decorate,decoration={brace,amplitude=5pt,raise=4pt},yshift=0pt, xshift=0.7cm]
(\b,\x) -- (\b,\z) node [black,midway,xshift=0.6cm] {\normalsize
$0.3$};
\draw [decorate,decoration={brace,amplitude=5pt,raise=4pt},yshift=0pt, xshift=0cm]
(\b,\x) -- (\b,\y) node [black,midway,xshift=0.6cm] {\normalsize
$0.7$};
\draw [decorate,decoration={brace,amplitude=5pt,raise=4pt},yshift=0pt, xshift=0cm]
(\b,\y) -- (\b,\z) node [black,midway,xshift=0.6cm] {\normalsize
$0.3$};

%%%%%%%%%%%%%%%%%%%%%%%%%%%%%%%%%%%%%%%%%%%%%%%%%%%%%%

\draw[->] (-1,\z-\e) -- (-1,\x+\e);
\node[left] at (-1,\x) {\normalsize most};
\node[left] at (-1,\x-2*\e) {\normalsize preferred};
\node[left] at (-1,\x-4*\e) {\normalsize according};
\node[left] at (-1,\x-6*\e) {\normalsize to $\succsim_R$};

\end{tikzpicture}